\documentclass[showpacs,doublecol,floatfix]{epl2}
\usepackage{epsfig}
\usepackage{graphics}
\usepackage{bm}
\usepackage[latin2]{inputenc}
\usepackage[T1]{fontenc}
\usepackage{array}

\usepackage{amssymb,amsmath}

\title{Exotic magnetic orders for high spin ultracold fermions}
\author{E. Szirmai\inst{1,2} \and  M. Lewenstein\inst{2,3}}

\institute{\inst{1} Research Institute for Solid State Physics and
Optics, H-1525 Budapest, P. O. Box 49, Hungary
 \\ \inst{2} ICFO-Institut de Ci{\`e}ncies Fot{\`o}niques,
Mediterranean Technology Park, E-08860 Castelldefels (Barcelona),
Spain
\\  \inst{3} ICREA-Instituci{\'o} Catalana de Recerca i Estudis Avan\c{c}ats,
Lluis Companys 23, E-08010 Barcelona, Spain}


\abstract{We study Hubbard models for ultracold bosonic or
fermionic atoms loaded into an optical lattice. The atoms carry a
high spin $F>1/2$, and  interact on site via strong repulsive Van
der Waals forces. Making  convenient rearrangements of the
interaction terms, and exploiting their symmetry properties, we
derive low energy effective models with nearest-neighbor
interactions, and their properties. We apply our method to $F=3/2$, and $5/2$
fermions on two-dimensional square lattice at quarter, and $1/6$ fillings, respectively, and
investigate mean-field equations for repulsive couplings. We find for $F=3/2$ fermions that the plaquette state
appearing in  the highly symmetric SU(4) case 
does not require fine tuning, and  is stable in an extended region of
the phase diagram. This phase competes with an SU(2) flux state,
that is always suppressed for repulsive interactions in absence of
external magnetic field. The SU(2) flux state has, however,
lower energy than the plaquette phase, and stabilizes in the
presence of weak applied magnetic field. 
For $F=5/2$ fermions a similar SU(2) plaquette phase is found to be the ground state 
without external magnetic field.}

\begin{document}

\maketitle

Ultracold atoms in optical lattices provide controllable quantum
many body systems that allow to mimic condensed matter~\cite{Lewenstein07,Bloch08}. They may in particular serve as
quantum simulators of various Hubbard models~\cite{Jaksch05},
including those that do not have condensed matter analogues.
Prominent examples include Hubbard models for bosons or fermions
with high spin $F$. Experimental progress in studies of high $F$
 Bose-Eistein condensates~\cite{StamperKurn99} and Fermi gases (cf.~\cite{Schneider08}) triggered a lot of interest in theoretical
 studies of such models. These studies go back to fundamental
 questions of large $N$ limit of SU($N$) Heisenberg-Hubbard model~\cite{Marston89}; they have continued more recently in the context of
 ultracold  atoms~\cite{Zhang01,Mishra02,Harada03}.  These papers discusses the
 interplay between the N\'eel, and valence bond solid (VBS), i.e.
 Peierls or plaquette ordering for antiferromagnetic systems.
 Several other exotic phase should be possible of earth alkali
 atoms (cf.~\cite{Honerkamp04,Cazalilla09,Gorshkov10,Hermele09}),
 where two orbital SU($N$) magnetism, and even chiral spin liquid
 states were predicted. Several authors predicted
 also a variety of novel, exotic phases from  effective (generalized Heisenberg)
 spin Hamiltonians, obtained from spinor Hubbard models (cf.~\cite{Imambekov03,Eckert07}). A lot of
 effort was devoted to the investigations of 1D and 1D ladder systems,  where
 quantum effects are even stronger~\cite{1D}. While for $F=1/2$
 Hubbard models  quantum fluctuations suppress
the conductor-Mott insulator transition~\cite{Lieb68}, this is not
the case for higher $F$, where dimer (Peierls) or valence bond
crystal (Haldane) order, and in ladders even plaquette order are
possible.

Fermi systems with $F=3/2$ were also intensively
studied~\cite{Wu}: first, because this is the simplest case beyond
$F=1/2$, second, because they can be realized with for instance
with ultracold $^{132}$Cs, $^{9}$Be, $^{135}$Ba, $^{137}$Ba, and
$^{201}$Hg (for an excellent review see Ref.~\cite{WuR}; Such
systems exhibit a generic SO(5) or isomorphically, Sp(4)
symmetry). In 1D there exist the quartetting phase, a four-fermion
counterpart of the Cooper pairing phase. In some situations,
counter intuitively  quantum fluctuations in spin-3/2 magnetic
systems are even stronger than those in spin-1/2 systems.

In this Letter we study the two-dimensional Hubbard model for $F=3/2$ fermions
with repulsive singlet and quintet interactions, and 
for $F=5/2$ fermions in a 2D plane of its 3D parameter space.
First, by rearranging the interaction terms and exploiting their
symmetry properties, we derive low energy effective Hamiltonians.
In contrast to the standard approaches (see for instance
~\cite{Imambekov03,Eckert07}), we do not use the spin
representation, but rather keep the description in terms of
fermionic operators. This allows us to formulate mean-field
theory, somewhat analogous to slave-boson method~\cite{Lee06}, and
show that the plaquette VBS state is stable in an extended region
of the phase diagram, in agreement with the  predictions of
Ref.~\cite{Wu,WuR}. In the presence of weak applied magnetic
field, however, the plaquette phase can be suppressed by an exotic
SU(2) flux state. Moreover, for $F=5/2$ fermionic atoms 
similar SU(2) plaquette phase can be the ground state of the system without 
external magnetic field.

Let us consider a system described by a Hamiltonian with
nearest-neighbor hopping $H_{kin}= -t\sum_{<i,j>}
c_{i,\alpha}^\dagger c_{j,\alpha}$ and strong on-site repulsive
interaction
\begin{equation}
 \label{eq:hint}
H_{int}= \sum_i V^{\alpha,\beta}_{\gamma,\delta} c_{i,\alpha}^\dagger
c_{i,\beta}^\dagger c_{i,\delta} c_{i,\gamma}.
\end{equation}
 $c_{i,\alpha}^\dagger$
($c_{i,\alpha}$) are the usual creation (annihilation) operators of
fermions with spin $\alpha$ at site $i$, and $t$ is the hopping
amplitude between the neighboring sites. Here, and in the
following automatic summation over repeated Greek indices is
assumed.
 The interactions depend on
the spin of the scattering particles \cite{Ho-Ohmi}:
\begin{equation}
\label{ham1}
V^{\alpha,\beta}_{\gamma,\delta}=\sum_{S=|F_1-F_2|}^{|F_1 + F_2|} g_S
\left[P_S\right]^{\alpha,\beta}_{\gamma,\delta},
\end{equation}
where $S$ is the total spin of the two scattering spin-$F_1$ and spin-$F_2$ particles.
 This means that the scattering processes can happen at different spin
 channels which are determined by the total spin of the scattering
 particles.  $P_S$ projects to the total spin-S subspace and $g_S$ is
 the coupling constant in the corresponding scattering channel.  Due
 to the on-site interaction the only contributing terms are either
 antisymmetric ($as$) or symmetric ($s$) for an exchange of the spin
 of the colliding particles depending on their fermionic or bosonic nature.
In the following we exploit this property of the on-site
 interaction which is also preserved for the effective strong repulsion model with nearest-neighbor interaction.

Starting from the fundamental relation between the $P_S$ projection operator and the product
of the $\mathbf{F}$ spin operators for two spin-$F$ fermions: 
$\left(2\mathbf{F}_1 \mathbf{F}_2\right)^l=\sum_S \left[ S(S+1) - 2 F(F+1) \right]^l P_S$,
the $P_S$ projector can be expressed as a degree of $2F$
polynomial of the  $\mathbf{F}_1 \mathbf{F}_2$ product:
\begin{equation}
\label{eq:s-p}
 P_{S} = \sum_{l=0}^{2F} a_{S,l} \left(\mathbf{F}_1 \mathbf{F}_2\right)^l
\end{equation}
for all $S=0,1,\ldots,2F$ and  with $\left(
\mathbf{F}_1 \mathbf{F}_2 \right)^0 \equiv\mathbf{E}$. $a_{S,l}$ are the coefficients of the
expansion. Note that for a given value of $S$ the $P_{S}$ projector is either symmetric or antisymmetric 
in the spin indices of the scattering particles.
 This expansion is usually applied in order to express the high-spin two-particle interaction with
effective multispin-exchange. In contrast, we will use the expansion of the projector operator in eqs.~\eqref{eq:hint} and ~\eqref{ham1}
in order to collect and treat adequately the two-particle interaction
terms that describe different spin exchange and spin flip processes.
In this case $\bf{F}$ denotes the three generators of the SU(2) Lie
algebra in the appropriate representation: for high-spin fermions or bosons they are the SU(2) generators
represented by the proper even/odd dimensional matrices. 
In case of pure boson or fermion system all processes take place only in the symmetric or antisymmetric
part of the total spin space, respectively. Therefore, the following decomposition can be used:
\begin{flalign}
\left(\mathbf{F}_1 \mathbf{F}_2\right)^l = & \left( \left( \mathbf{F}_1 \mathbf{F}_2 \right)^l \right)^{(as)} +
\left( \left( \mathbf{F}_1 \mathbf{F}_2 \right)^l \right)^{(s)},
\end{flalign}
and the symmetric and antisymmetric
projectors can be expressed as follows:
\begin{subequations}
\label{psdec}
\begin{flalign}
P_S^{(as)}= & \sum_{l=0}^{N_{(as)}-1} b_{S,l} \left( \left(
\mathbf{F}_1 \mathbf{F}_2 \right)^l \right)^{(as)}, \\ P_S^{(s)}= &
\sum_{l=0}^{N_{(s)}-1} c_{S,l} \left( \left( \mathbf{F}_1
\mathbf{F}_2 \right)^l \right)^{(s)}.
\end{flalign}
\end{subequations}
Here $N_{(as)}$ and $N_{(s)}$ denotes the number of antisymmetric
and symmetric subspaces of the total spin space. The
antisymmetric and symmetric part of an operator $A$ can be constructed
by the exchange of two spin indices: $ \left[A^{(as)}\right]^{\alpha,\beta}_{\gamma,\delta} = 
 A^{\alpha,\beta}_{\gamma,\delta} - A^{\alpha,\beta}_{\delta,\gamma}$, and
$\left[A^{(s)}\right]^{\alpha,\beta}_{\gamma,\delta} = 
 A^{\alpha,\beta}_{\gamma,\delta} + A^{\alpha,\beta}_{\delta,\gamma}$.
It is obvious that the above decomposition leads to the polynomials Eq.~\eqref{psdec} 
having significantly smaller degree than Eq.~\eqref{eq:s-p}.
$N_{(as)}-1$ or $N_{(s)}-1$, respectively, determines the minimum degree of the polynomial of the product $\mathbf{F}_1 \mathbf{F}_2$
which is equivalent to the interaction Eq.~\eqref{ham1}.

Now let us apply the above procedure to a 2 dimensional $F=3/2$ fermion system.
In this case the interaction has to
 be antisymmetric therefore the only contributing terms are the total
 spin-0 (singlet) and the spin-2 (quintet) scatterings:
\begin{equation}
\label{int-1}
V^{\alpha,\beta}_{\gamma,\delta}=\sum_{S=0}^3 g_S
\left[P^{(as)}_S\right]^{\alpha,\beta}_{\gamma,\delta} =g_0
\left[P_0\right]^{\alpha,\beta}_{\gamma,\delta} + g_2
\left[P_2\right]^{\alpha,\beta}_{\gamma,\delta}.
\end{equation}
At quarter filling there is only one particle on each site (i) on the average in general, 
and (ii) exactly if the on-site repulsion tends to infinity. Therefore for strong
repulsion, the hopping can be considered as a perturbation, and
 the system can be described by an effective Hamiltonian with nearest-neighbor
interaction. The effective model
based on perturbation theory up to second (leading) order in the
hopping $t$ is the following:
\begin{equation}
 H_{eff}= \sum_{<i,j>}
 \tilde{V}^{\alpha,\beta}_{\gamma,\delta} c_{i,\alpha}^\dagger
 c_{j,\beta}^\dagger c_{j,\delta} c_{i,\gamma},
\end{equation}
where $\tilde{V}^{\alpha,\beta}_{\gamma,\delta}= \sum_{S} G_S
\left[P_S^{(as,s)}\right]^{\alpha,\beta}_{\gamma,\delta}$ gives the energy shift 
to the on-site energies due to the weak nearest-neighbor
hopping. $G_S=
-4t^2/g_S$, ($S=0,2$) is the new coupling constant in the spin-$S$ scattering channel. Since the effective model preserves the symmetry of the on-site
interaction, it remains antisymmetric for an exchange
of two spin indices. Now the
components of the $\bf{F}$ vector are the well known $4\times 4$ spin matrices and Eq.~\eqref{psdec} has the following form:
$\mathbf{E}^{(as)} = P_0 + P_2$, and $\left(\mathbf{F}_1 \mathbf{F}_2 \right)^{(as)} = -15 P_0/4
-3 P_2/4$. The effective Hamiltonian has the form
\begin{equation}
H_{int} = a_n \sum_{<i,j>} \mathbf{E}_{i,j}^{(as)} + a_s \sum_{<i,j>}
\left(\mathbf{F}_1 \mathbf{F}_2 \right)^{(as)}_{i,j},
\end{equation}
where $a_n=(5 G_2 - G_0)/4$, and $a_s=(G_2 - G_0)/3$. The two-particle nearest-neighbor interaction terms are:
$\mathbf{E}_{i,j}^{(as)}= 
c_{i,\alpha}^\dagger c_{j,\beta}^\dagger c_{j,\delta} c_{i,\gamma}
\left[\mathbf{E}^{(as)}\right]^{\alpha,\beta}_{\gamma,\delta}$, and $(\mathbf{F}_1 \mathbf{F}_2 )^{(as)} = 
 c_{i,\alpha}^\dagger
c_{j,\beta}^\dagger c_{j,\delta} c_{i,\gamma} \left[
  (\mathbf{F}_1 \mathbf{F}_2 )^{(as)}
  \right]^{\alpha,\beta}_{\gamma,\delta}$,
and their explicit spin dependence is:
 $\left[\mathbf{E}^{(as)}\right]^{\alpha,\beta}_{\gamma,\delta} = 
 \delta_{\alpha,\gamma}\delta_{\beta,\delta} -
 \delta_{\alpha,\delta}\delta_{\beta,\gamma}$, and $\left[
   (\mathbf{F}_1 \mathbf{F}_2 )^{(as)}
   \right]^{\alpha,\beta}_{\gamma,\delta} =  \left[\mathbf{F}_1
   \right]_{\alpha,\gamma} \left[\mathbf{F}_2 \right]_{\beta,\delta} -
 \left[\mathbf{F}_1 \right]_{\alpha,\delta} \left[\mathbf{F}_2
   \right]_{\beta,\gamma}$.
After straightforward
calculations one arrives to the following form of the effective Hamiltonian:
\begin{multline}
\label{effham}
 H_{eff} = V^{(0)} + \sum_{<i,j>} \bigg[ a_n \Big( n_i n_j +
   \chi_{i,j}^\dagger \chi_{i,j} - n_i \Big) \\ + a_s \Big(
   \mathbf{S}_i \mathbf{S}_j + \mathbf{J}_{i,j}^\dagger
   \mathbf{J}_{i,j} -\frac{15}{4} n_i \Big) \bigg],
\end{multline}
where $n_i =  c_{i,\alpha}^\dagger c_{i,\alpha}$, and
$\mathbf{S}_i =  c_{i,\alpha}^\dagger
\mathbf{F}_{\alpha,\beta} c_{i,\beta}$ are the usual particle number
and spin operators on site $i$, and
\begin{subequations}
 \begin{flalign}
\chi_{i,j} = &  c_{i,\alpha}^\dagger c_{j,\alpha},
\\ \label{sdef} \mathbf{J}_{i,j} = &
c_{i,\alpha}^\dagger \mathbf{F}_{\alpha,\beta} c_{j,\beta}
 \end{flalign}
\end{subequations}
are introduced for the U(1), and SU(2) nearest-neighbor link
operators, respectively. Note, that in general the SU(2) link
operators do not satisfy the spin commutation
relations, however, they clearly are related to the bond-centered spin. 
The competition between the spin and particle fluctuations can be
controlled by tuning of $a_n$ and $a_s$. The effective Hamiltonian
\eqref{effham} can be applied for less than quarter filled system
too, provided the kinetic term is added to the Hamiltonian
\eqref{effham}. $V^{(0)}$ contains the on-site energies and shifts
the ground state energy only, so we do not consider its contribution. 

In the following we study the possible phases of the quarter filled system with the
constraint $\sum_\alpha c_{i,\alpha}^\dagger c_{i,\alpha} = 1$ --- only single
occupied sites are allowed due to the strong on-site repulsion. Due to this
local constraint the Hamiltonian is invariant under a rotation of the phase
of the fermions at each sites. This means that the Lagrangian of the system
$\mathcal{L}= \sum_i c_{i,\sigma}^\dagger \partial_\tau c_{i,\sigma} + H$
is invariant under the U(1) gauge transformation $c_{i,\sigma}\rightarrow
c_{i,\sigma} e^{i\phi_i}$ reflecting the local constraint for the particle
number.

Considering the Hamiltonian \eqref{effham}
the terms containing $n_i$ do not give contribution up to an irrelevant constant at quarter filling and the
remaining 4-fermion terms can be decoupled via a mean-field treatment by introducing the expectation
values of the link operators $\left<\chi_{i,j}\right>$ and $\left<\mathbf{J}_{i,j}\right>$, and the spin
operator $\left<\mathbf{S}_i\right>$. Now the mean-field Hamiltonian is:
\begin{equation}
\label{mfham}
 H^{MF}= \textstyle{\sum_{<i,j>}} H_{i,j},
\end{equation}
with
\begin{multline*}
 H_{i,j} =  a_n \Big(
   \left<\chi_{j,i}\right> c_{i,\alpha}^\dagger c_{j,\alpha} +
   \left<\chi_{i,j}\right> c_{j,\alpha}^\dagger c_{i,\alpha} -
   |\left<\chi_{i,j}\right> |^2 \Big) \\ + a_s
   \Big( \left< \mathbf{J}_{j,i} \right>
   c_{i,\alpha}^\dagger \mathbf{F}_{\alpha,\beta} c_{j,\beta} + \left<
   \mathbf{J}_{i,j} \right> c_{j,\alpha}^\dagger
   \mathbf{F}_{\alpha,\beta} c_{i,\beta} - | \left<
   \mathbf{J}_{i,j} \right> |^2 \\ + \left< \mathbf{S}_i\right>
c_{j,\alpha}^\dagger \mathbf{F}_{\alpha,\beta} c_{j,\beta} + \left< \mathbf{S}_j \right> c_{i,\alpha}^\dagger
\mathbf{F}_{\alpha,\beta} c_{i,\beta} - \left<
   \mathbf{S_i} \right> \left< \mathbf{S}_j \right> \Big) .
\end{multline*}
Note that the mean-field Lagrangian also has to remain invariant
under the gauge transformation mentioned above. Thus, the link
variables must transform as $\left<A_{i,j}\right> \rightarrow
\left< A_{i,j} \right> e^{-i(\phi_j-\phi_i)}$.

\begin{figure}
\centering
\includegraphics[scale=0.31]{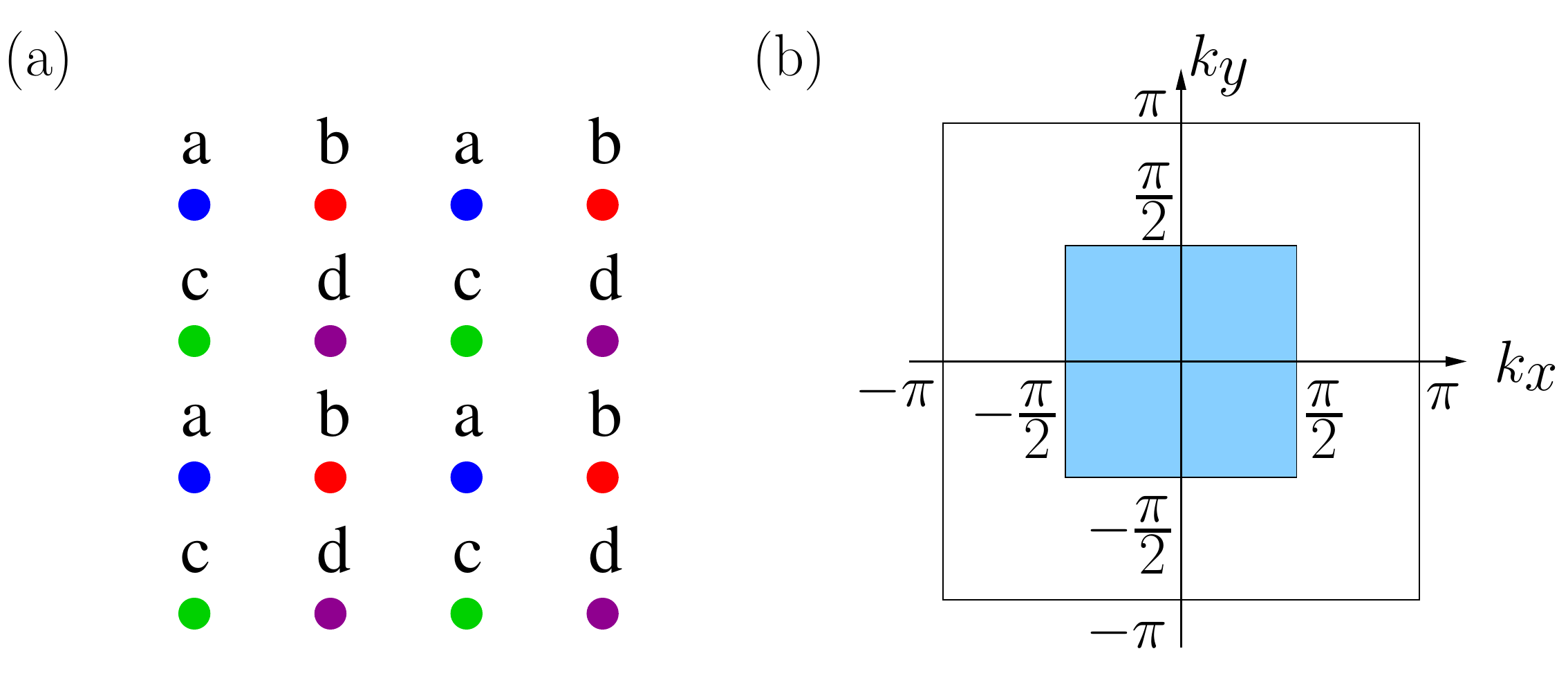}
\caption{(Color online) The lattice was split into 4 sublattices (a) and due to this splitting the Brillouin zone shrank. (b) The shadow
area depicts the reduced Brillouin zone.}
\label{fig:sublattice+BZ}
\end{figure}

The expectation values of the spin and link operators were
determined self-consistently. Anticipating the appearance of a
plaquette phase similar to the ground state of the
system for $G_0=G_2$, it is reasonable to split the lattice into 4
sublattices (see Fig. \ref{fig:sublattice+BZ}) leading to the
shrinking of the Brillouin zone to the quarter of its
original size. We assume different values for the order
parameters of the different sublattices and of the alternating
links as the only space-dependence of them. 

\begin{figure}
\centering
\includegraphics[scale=0.29]{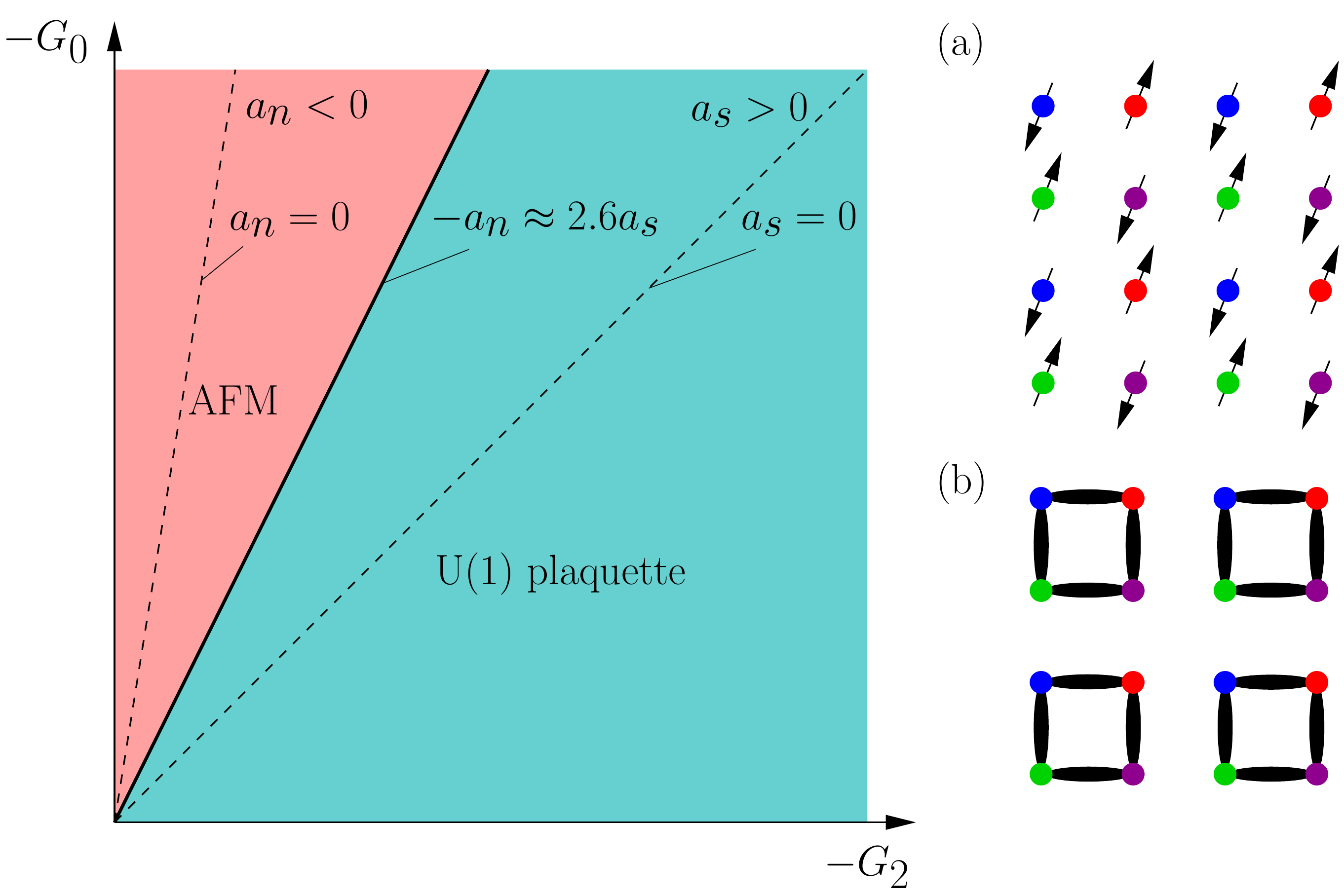}
\caption{(Color online) The phase diagram of the spin-3/2 fermion system with strong
on-site repulsion on 2D square lattice at quarter filling and the corresponding configurations: (a) N\' eel order (AFM), and 
(b) U(1) plaquette phase.
}
\label{fig:phase-diag}
\end{figure}

We have found the following gauge non-equivalent states with the following nonzero averages to be the
 solutions of the self-consistent equations: 
(a) N\' eel order: $\left< \mathbf{S}_j \right>$,
(b) U(1) plaquette order: $\left<\chi_{i,j}\right>$, 
and (c) SU(2) plaquette and dimer order: $\left< \mathbf{S}_j \right>$, 
$\left<\chi_{i,j}\right>$, $\left< \mathbf{J}_{j,i} \right>$.
In the figures 
the nonzero expectation values are denoted by the following way: $\left< \mathbf{S}_j \right>$: black arrow, $\left<\chi_{i,j}\right>$: 
black stripe, and $\left< \mathbf{J}_{j,i} \right>$: light blue stripe with arrow. 
The phase diagram of the system is shown in
Fig.~\ref{fig:phase-diag}. If the effective interaction of the
singlet channel is significantly stronger than that of the
quintet channel, the dominant order is purely antiferromagnetic
without any bond order. 
For $a_n<0$ and $a_s>0$ the spin
and particle order compete with each other. In ref.~\cite{Wu} a
magnetically ordered dimer phase was suggested to appear in this
regime, however, we did not find any similar state to be the solution of the self-consistent equations: 
when the antiferromagnetic order of the N\' eel phase
is destroyed, plaquette order appears. The phase boundary is
around $G_0\approx 1.9 G_2$ or equivalently $-a_n\approx 2.6 a_s$.
In the U(1) plaquette phase the coupling constant $a_n$ always dominates the interaction 
 independently of the
sign of $a_s$. The nonzero U(1) links form boxes as
shown in Fig.~\ref{fig:phase-diag} (b). One can define the
U(1) plaquette variable as $\Pi =\chi_{i,j}\chi_{j,k}\chi_{k,l}\chi_{l,i}$,
where $i$, $j$, $k$ and $l$ denote the sites of an elementary
plaquette of the square lattice, and $\chi$ is defined for 
nearest-neighbors only. The U(1) flux $\Phi$ is defined by the phase of
the plaquette. The plaquette variable and therefore the flux are invariant
under the U(1) gauge transformations mentioned above. We have found two different
gauge-non-equivalent states in the plaquette phase labeled by
$\Phi=0$ and $\Phi=\pi$, respectively, and both states have the
same energy.
Note that our results are in good agreement 
with earlier results for the special SU(4) symmetric case ($G_0=G_2$), 
where similar box state was predicted with zero flux \cite{Mishra02}. We have found that this box state
does not need fine tuning, it is the ground state in an extended, experimentally reachable
 region of the parameter space.

\begin{figure}
\centering
\includegraphics[scale=0.29]{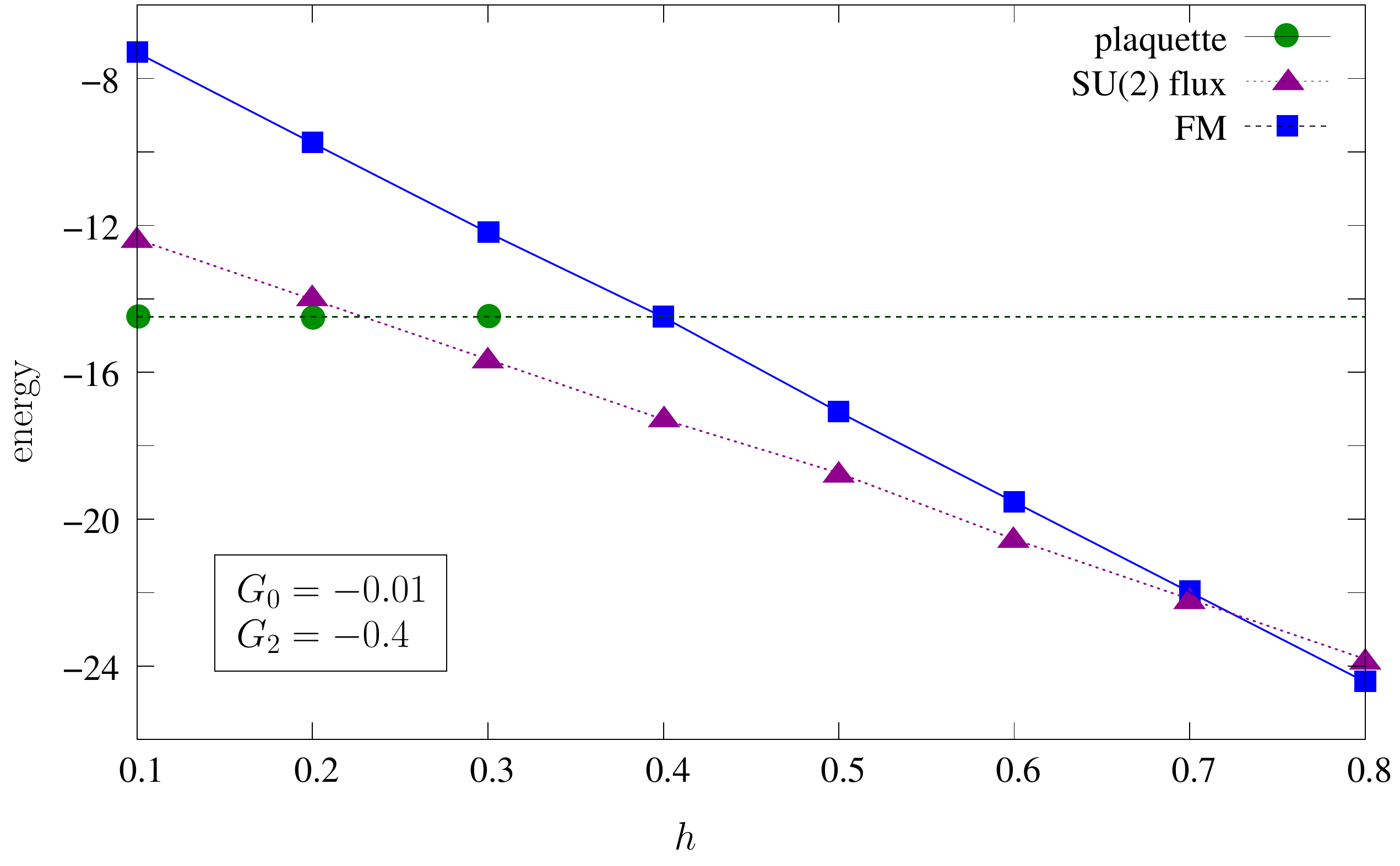}
\caption{(Color online) The magnetic field dependence of the energy of the
different states in units of $t$ (the
lines are only guides of the eyes), and the configurations in the SU(2) plaquette (a) and in the SU(2) dimer (b) phase.
} \label{fig:h-dependence}
\end{figure}

In the parameter region
$|G_0|< 1.9 |G_2|$, and for $a_s\neq 0$ we have found two other solutions on the top of the ground state
having 10-15\% higher energy: the SU(2)
dimer phase and the SU(2) plaquette phase, where the latter corresponds
 to two gauge-non-equivalent states with different fluxes. Both the dimer and the plaquette phases have the same energy.
In these states, in addition to weak ferromagnetic order ($\left<\mathbf{S}_i\right> < 3/2$ and are equal for all of the 4 sublattices),
both types of the link operators $\chi$ and $\mathbf{J}$ have
nonzero expectation values as shown in
Fig.~\ref{fig:h-dependence} (a), and (b).
In the SU(2) plaquette phase the link
operators with nonzero expectation values form plaquettes.
 These states are completely new RVS orders, therefore we make some notes about 
their basic properties and comment on their naming.
Both states violate spin-rotation invariance, and the SU(2) dimer state --- contrary to the SU(2) plaquette
phases --- preserves the translational invariance by one lattice
site in one spatial dimension. 
It is clear that $\mathbf{J}$ is not a member of SU(2) therefore it is
reasonable to ask why do we use the terms SU(2) plaquette, flux or
dimer for the states where the expectation value of $\mathbf{J}$
is nonzero? To answer this question let us consider the mean-field
Hamiltonian Eq.~\eqref{mfham}. The non-local part of the
one-particle excitations appears in the Hamiltonian as
\begin{equation}
\label{eq:nonloc}
 \Big( a_n  \left<\chi_{j,i}\right>\delta_{\alpha,\beta} + a_s \left<\mathbf{J}_{i,j}\right>
\mathbf{F}_{\alpha,\beta} \Big) c_{i,\alpha}^\dagger c_{j,\beta} + H.c.
\end{equation}
From this form it can be read that the excitations consist two branches with two different symmetries:
$\left<\chi_{j,i}\right>$ relates to the U(1) excitations, while $\left<\mathbf{J}_{i,j}\right>\mathbf{F}$ to the SU(2) excitations.
In order to define the SU(2) flux let us introduce the new link parameter according to Eq.~\eqref{eq:nonloc}:
$U_{i,j} = \left<\mathbf{J}_{i,j}\right> \mathbf{F}$,
with the usual inner product of the vectors in the 3 dimensional
space of the generators $\mathbf{F}$. $U_{i,j}$ is a member of
SU(2) and a $4\times 4$ matrix for $F=3/2$ fermionic atoms and the same holds for the
SU(2) plaquette variable $\Pi^{SU(2)} = U_{i,j} U_{j,k} U_{k,l} U_{l,i}$.  The
flux $\mathbf{\Phi}$ passing through the plaquette defined by the
form: $\Pi^{SU(2)}=e^{i\mathbf{\Phi}\mathbf{F}}$. In order to
determine the ground state it is worth to express the mean-field
Hamiltonian with the $\mathbf{J}_{i,j}$ operators, while the
excitations and the SU(2) flux can be expressed with $U_{i,j}$.
Note that the SU(2) plaquette $\Pi^{SU(2)}$ is also invariant
under the U(1) gauge transformation defined above:
$c_{i,\sigma}  \rightarrow c_{i,\sigma} e^{i\phi_i}$, $\left<\chi_{i,j} \right>  
\rightarrow \left< \chi_{i,j} \right>
e^{i(\phi_j-\phi_i)}$, and $U_{i,j}  \rightarrow U_{i,j} e^{i(\phi_j-\phi_i)}$.
Considering the definition of $U_{i,j}$, the last relation is obviously equivalent to the transformation
$\mathbf{J}_{i,j} \rightarrow \mathbf{J}_{i,j}e^{i(\phi_j-\phi_i)}$.

The SU(2) phases can patently claim to great interest, but they are
suppressed by the U(1) plaquette state. Nevertheless, since the
SU(2) flux, as well as the SU(2) dimer order coexist with
ferromagnetic order, it can be expected that weak magnetic field
does not destroy the SU(2) order, but it can stabilize that. To
check this let us investigate the energy of these states in the
presence of external magnetic field $h$ taken into account as a
Zeeman term in the Hamiltonian: 
\begin{equation}
H^{h}= H^{MF} + h\sum_{i} \mathbf{S}_i.
\end{equation}
Note that for strong magnetic field the quadratic Zeeman term can become important.
At this point let us suppose that this term can be neglected and we will 
check the validity of this assumption at the end 
of the calculations.

The magnetic field dependence of the energy of the SU(2) flux
state compared to the U(1) plaquette state and to the ferromagnetic
state is shown in Fig.~\ref{fig:h-dependence} for a typical value
of the couplings in units of the nearest-neighbor hopping. The
U(1) plaquette phase remains the ground state in the presence of nonzero, but
very small  magnetic fields. The SU(2) plaquette state, as
well as SU(2) dimer order, have the lowest energy for higher value of the applied magnetic
field $h$, and are the ground state of the system
in an extended region of the phase diagram. Both
the SU(2) dimer and the plaquette states either 0 or $\pi$ flux have the same energy. 
In the presence of even stronger magnetic fields ferromagnetic order suppresses any other order in the system.
Now, let us check the validity of our assumption of neglecting the quadratic Zeeman coupling. 
The linear Zeeman energy (in $\hbar$ unit) is given by the Lamour frequency:
$\omega_L = g_F \mu_B B$,
and the quadratic Zeeman energy is:
$\omega_q = \frac{\omega_L^2}{\omega_{hf}}$.
Here $\omega_{hf}$ is the hyperfine splitting energy, $g_F$ is the 
gyromagnetic factor and $\mu_B$ is the Bohr magneton. If $\frac{\omega_L}{\omega_q}\gg 1$ 
or equivalently
$\frac{\omega_{hf}}{\omega_L}\gg 1$,
the quadratic Zeeman effect can be neglected.
We measure the magnetic field in units of the hopping 
parameter $t$. In optical lattices
$t=\omega_R \frac{2}{\pi} \xi^3 e^{-2\xi^2}$,
where $\xi=(V_0/\omega_R)^{1/4}$, $V_0$ is the potential depth, and $\omega_R$ 
is the recoil energy. $t$ has a maximum at $V_0\approx \omega_R$, where $t\sim \omega_R$. $\omega_R$ is 
typically in the order of $1-100$ kHz (however, $\omega_R /2\pi \sim 400.98$ kHz for $^9$Be). We have found that 
the SU(2) flux state has the lowest energy if the linear Zeeman energy is around $\omega_L\sim 0.1-1t$ 
which means 0.1-100 kHz. The hyperfine frequency is in the order of 1-10 GHz, so 
$\omega_{hf}/\omega_L$ is in the order of $10^4-10^7$. Therefore, the quadratic Zeeman effect can 
be neglected compared to the linear Zeeman term for magnetic fields which are sufficient to stabilize the SU(2) phases. There are some atoms which could be promising candidates for realizing experimentally spin-3/2 fermion systems and have much less recoil energy than the above mentioned $^6$Be, namely $1-10$ kHz which in our case corresponds to very small magnetic field: $10^{-5}-10^{-4}$ G. With these atoms the experimental realization of the SU(2) orders demands quite strong magnetic shielding.

\begin{figure}
\centering
\includegraphics[scale=0.29]{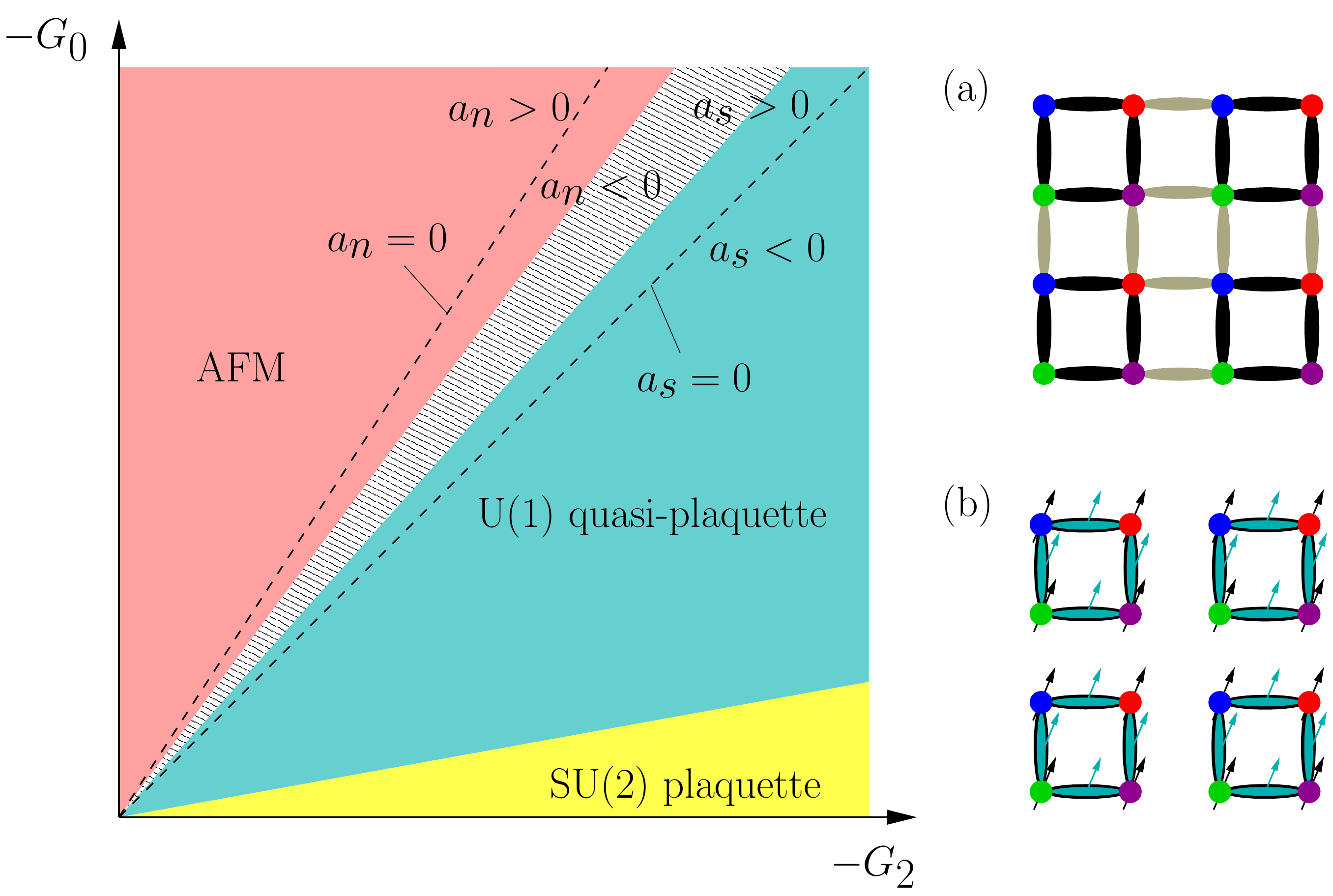}
\caption{(Color online) The phase diagram of the $F=5/2$ fermion system with strong
on-site repulsion on 2D square lattice at 1/6 filling and the configurations in the U(1) quasi-plaquette phase (a), and in the SU(2) plaquette phase (b). 
}
\label{fig:5/2}
\end{figure}

Similar analysis can be easily made for $F=5/2$ fermions at 1/6 filling for the 
special values of the coupling constants
$G_4\approx (-7 G_0 + 10 G_2)/3$. $G_4$ is the coupling of the
interaction with 9-fold spin multiplicity that appears in the Hamiltonian 
Eq. \eqref{int-1} for spin-5/2 system
in addition to the singlet ($G_0$) and quintet ($G_2$) scatterings. In the plane of the parameter space defined by $G_4=(-7 G_0 + 10 G_2)/3$,
 the structure of the Hamiltonian is exactly the same as Eq.~\eqref{mfham},
there is no term containing higher order of the product $\mathbf{F}_1 \mathbf{F}_2$, and
the couplings take the values: $a_n=(-23 G_0 + 35 G_2)/12$ and $a_s=(-G_0 + G_2)/3$. Note, that now the components of the $\mathbf{F}$ vectors are the 3 generators 
of the SU(2) Lie algebra in $6\times 6$ representation. During
 the calculations we have used the same 4-sublattice 
ansatz as in case of the $F=3/2$ fermions, because the more suitable 6-sublattice ansatz does not respect the symmetries of the original model on square lattice. Similarly we have determined 
the solutions of the self-consistent equations for the $\left< \mathbf{S}_i\right>$, $\left< \chi_{i,j}\right> $ and $\left< \mathbf{J}_{i,j} \right>$ 
expectation values and the corresponding energies.
 The ground states of this system
are shown in Fig.~\ref{fig:5/2}. 
In case of dominant singlet scatterings the
ground state is purely antiferromagnetic, at least while $a_n>0$.
A weak negative $a_n$ seems to lead to an instability in the
system, but we could not find any stable solution of the self-consistent equations in this narrow
region. Further decreasing $|G_0|$, a quasi-plaquette phase
appears. In this phase the expectation value of the U(1) link
operator $\left<\chi_{i,j}\right>$ is non-zero everywhere, but
stronger and weaker links alternate forming a weak plaquette
structure. The flux passing through the plaquette is zero and
there is no spin order in this phase. For even weaker singlet
coupling (increasing the value of $|a_s|$) weak ferromagnetic
order appears in addition to the plaquette order (the SU(2) plaquette phase in Fig.~\ref{fig:5/2}). Here the
plaquettes are formed not only by the alternating zero and
non-zero U(1) link operators $\left<\chi_{i,j}\right>$, but the
SU(2) operators $\left< \mathbf{J}_{i,j} \right>$, too. The flux
passing through the plaquettes remains zero. This means that while for $F=3/2$ fermions the SU(2) plaquette phase
 can be the ground state of the system only by applying external magnetic filed, for $F=5/2$ fermions a similar SU(2) plaquette order (with zero flux) 
is the ground state of the system for weak singlet couplings. It is difficult to determine precisely the phase border between the quasi-plaquette and the SU(2) plaquette phases. This is due to the fact that the two orders start to compete around $2 G_0\approx G_2$: both are stable solution of the self-consistent equations but approximately with the same energy. The energy difference between the two phases increases slowly for decreasing $|G_0|$, and around $3 G_0 \approx G_2$ reaches the $4-5\%$. Note, that in the
same parameter regime we have found another stable solution: a plaquette
phase with $\pi$ flux and with stronger ferromagnetic order,
however $\left< \mathbf{S}_i \right>$ remains smaller than 5/2. In
this state the SU(2) order
 parameter dominates the link variables: the value of $\left< \mathbf{J}_{i,j} \right> $  is twice than that of
 the corresponding $\left< \chi_{i,j}\right>$. The energy of this spin ordered SU(2)
 plaquette state with $\pi$ flux is higher by
about 5\% than the one with zero flux.

Finally let us discuss the validity of our mean-field results. In two dimensions the mean-field 
solutions are expected to provide qualitatively reliable results. However, fluctuations around the 
mean-field results can become relevant, especially in the following two cases:
when the system is close to the phase boundary of a continuous transition, the correlation 
length becomes very large and fluctuations cannot be neglected. The present work focuses 
on the qualitative description of possible exotic states of matter. The states discussed above are the 
ground state of the system in an extended region of the parameter space and far enough from 
the phase boundaries expectedly they can be realized experimentally. 
Fluctuations could also be important if the mean-field solution is degenerate and the states in the degenerate 
subspace are not separated by energy barrier. In our case e.g. the SU(2) dimer and plaquette 
states have the same energy but they are separated by energy barrier: one cannot arrive from 
one to the other with continuous (link by link) deformation without increasing the energy. 
These types of mean-field solutions are expectedly not effected by small fluctuations.

To summarize, we have used a decomposition of the total spin space into
its symmetric and antisymmetric part with respect to the exchange
of two spin indices of the high spin scattering particles. This
decomposition was used for strongly repulsive systems to derive 
effective low energy Hamiltonians. This task was achieved remaining
within the two-particle representation. The main advantage of the
treatment is that it does not require to introduce complicated
effective multiparticle/multispin interactions, but relies  only
on rearrangements of the usual two-particle interactions. The
effectiveness of the treatment does not depend on the statistics
of the considered particles, and it allows us to
 identify the different processes in the spin channel within the concept of site and bond spin.
 Applying this method to $F=3/2$ fermions, we have determined the
ground state phase diagram of the system on mean-field level to
complete the earlier results known for some regimes of the
couplings. We have found that the VBS state is the ground state in an
extended region of the phase diagram,
while in the presence of weak magnetic field an exotic
SU(2) flux state has the lowest energy. We have also made some similar calculations for
$F=5/2$ fermions in the plane determined by the condition $G_4=(-7
G_0 + 10 G_2)/3$ in the 3-dimensional parameter space of the coupling
constants. 
We have found that novel, exotic SU(2) plaquette phase, similar to one predicted for spin-3/2 fermion system 
in the presence of external magnetic field, can be the ground state of the system with zero flux, without applied magnetic field.

\acknowledgments
This work was funding by the Spanish MEC projects TOQATA (FIS2008-00784),
QOIT (Consolider Ingenio 2010), ERC Grant QUAGATUA, EU STREP NAMEQUAM, and
partly (E. Sz.) by the Hungarian Research Fund (OTKA) under Grant No. 68340.


\begin{thebibliography}{0}

\bibitem{Lewenstein07} \Name{M. Lewenstein {\it et al.}}
\REVIEW{Adv. Phys.}{56}{2007}{243}.

\bibitem{Bloch08}
\Name{I. Bloch, J. Dalibard, \and W. Zwerger} 
\REVIEW{Rev. Mod. Phys.}{80}{2008}{885}.

\bibitem{Jaksch05} \Name{D. Jaksch \and P. Zoller} 
\REVIEW{Ann. Phys. (N.Y.)}{315}{2003}{52}.

\bibitem{StamperKurn99} \Name{D. Stamper-Kurn \and W. Ketterle} \REVIEW{Proceedings of Les Houches 1999 Summer School, Session
LXXII}{}{1999}{}.

\bibitem{Schneider08} \Name{U. Schneider {\it et al.}}
\REVIEW{Science}{322}{2008}{1520};
\Name{J\"ordens {\it et al.}}  \REVIEW{Nature}{55}{2008}{204};
\Name{T. Esslinger} \REVIEW{arXiv:1007.0012}{}{2010}{}.

\bibitem{Marston89} \Name{J.B. Marston \and I. Affleck} 
\REVIEW{Phys. Rev. B}{39}{1989}{11538}.

\bibitem{Zhang01} \Name{G.-M. Zhang \and S.-Q. Shen} 
\REVIEW{Phys. Rev. Lett.}{87}{2001}{157201}.

\bibitem{Mishra02} \Name{A. Mishra, M. Ma, \and F.-C. Zhang} 
\REVIEW{Phys. Rev. B}{65}{2002}{214411}.



\bibitem{Harada03} \Name{K. Harada, N. Kawashima, \and M.
Troyer} 
\REVIEW{Phys. Rev. Lett.}{90}{2003}{117203}.

\bibitem{Honerkamp04} \Name{C. Honerkamp \and W. Hofstetter} 
\REVIEW{Phys. Rev. Lett.}{92}{2004}{170403}.

\bibitem{Cazalilla09} \Name{M.A. Cazalilla, A.F. Ho \and M. Ueda} 
\REVIEW{New J. Phys.}{11}{2009}{103033}.

\bibitem{Hermele09} \Name{M. Hermele, V. Gurarie \and A.M. Rey} 
\REVIEW{Phys. Rev. Lett.}{103}{2009}{135301}.


\bibitem{Gorshkov10} \Name{A.V. Gorshkov {\it et al.}}
\REVIEW{Nature Physics}{6}{2010}{289}.


\bibitem{Imambekov03}   \Name{A. Imambekov, M. Lukin, \and E. Demler} 
\REVIEW{Phys. Rev. A}{68}{2003}{063602}.


\bibitem{Eckert07} \Name{K. Eckert {\it et al.}} 
\REVIEW{New J. Phys.}{9}{2007}{133}.


\bibitem{1D} \Name{R. Assaraf {\it et al.}} 
\REVIEW{Phys. Rev. Lett.}{93}{2004}{016407};
\Name{S. Chen {\it et al.}} 
\REVIEW{Phys. Rev. B}{72}{2005}{214428};
\Name{P. Lecheminant \and K.
Totsuka} 
\REVIEW{Phys. Rev. B}{71}{2005}{020407};
\Name{K. Buchta {\it et
al.}} 
\REVIEW{Phys. Rev. B}{75}{2007}{155108};
\Name{E. Szirmai, \"O. Legeza, \and J.
S\'olyom} 
\REVIEW{Phys. Rev. B}{77}{2008}{045106};
 \Name{H. Nonne  {\it et al.}}
\REVIEW{Phys. Rev B}{81}{2010}{020408R};
\Name{C. Wu} 
\REVIEW{Phys. Rev. Lett.}{95}{2005}{266404}.

\bibitem{Lieb68} \Name{L.H. Lieb \and F.Y. Wu} 
\REVIEW{Phys. Rev. Lett.}{20}{1968}{1445}.

\bibitem{Wu} \Name{C. Wu, J.-P. Hu, \and S.-C. Zhang} 
\REVIEW{Phys. Rev. Lett}{91}{2003}{186402};
\Name{C.Xu \and C. Wu}
\REVIEW{Phys. Rev. B}{77}{2008}{134449};
\Name{K. Rodr\'iguez {\it et al.}}
\REVIEW{Phys. Rev. Lett.}{105}{2010}{050402};

\bibitem{WuR}  \Name{C. Wu}
\REVIEW{Mod. Phys. Lett. B}{20}{2006}{1707}.


\bibitem{Lee06} \Name{P.A. Lee, N. Nagaosa \and X.-G. Wen} 
 \REVIEW{Rev. Mod. Phys.}{78}{2006}{17}.

\bibitem{Ho-Ohmi} \Name{T. L. Ho} \REVIEW{Phys. Rev. Lett.}{81}{1998}{742};
\Name{T. Ohmi, \and K. Machida} \REVIEW{J. Phys. Soc. Jpn.}{67}{1998}{1822}.


\end{thebibliography}
\end{document}